\documentclass[12pt]{article}
\usepackage{graphicx}
\large

\title{A quantum mechanical model of "dark matter"}

\vspace{0.5cm}

\author{ V.V. Belokurov$^{1,2}$ and E.T. Shavgulidze$^{1}$    \\
\\    {\em 1. Lomonosov Moscow State University, Russia }
\\    {\em 2. Institute for Nuclear Research of Russian Academy of Sciences, Russia }
\\ {  vvbelokurov@yandex.ru ; shavgulidze@bk.ru}}

\date{ \ \ \  }
\begin{document}
\maketitle

\begin{abstract}

The role of singular solutions in some simple quantum mechanical models is studied.
The space of the states of two-dimensional quantum harmonic oscillator is shown to be separated into sets of states with different properties.
\end{abstract}

\vspace{2cm}

\textbf{1. One-dimensional quantum oscillator with a singular potential.}

In our paper \cite{(BSh)} it was shown that paths with singularities are very important for understanding the functional integrals structure. Here we study the role of singular solutions in forming sophisticated structures of the spaces of states in some simple quantum mechanical models.
In particular, the space of states of two-dimensional quantum harmonic oscillator turned out to be mutually disjoint  clusters.

To make the effect we discuss in this paper more transparent, first let us consider the quantum oscillator with
the Hamiltonian
 \begin{equation}
   \label{1}
   H=\frac{1}{2}\left( -\frac{d^{2}}{dx^{2}}+x^{2}+2x^{-2}\right)\,.
\end{equation}

The wave functions
regular at $x=0$ are well known (see, e.g. \cite{(Land)},\cite{(Per)}). They are expressed in terms of degenerate hypergeometric functions and form a basis in the Hilbert space of functions on the semi-axis $ 0< x < +\infty$ bounded at $x=0\,.$

However, there is another class of solutions. To find them we represent the Hamiltonian (\ref{1}) as
\begin{equation}
   \label{2}
H=a^{+}_{\alpha}\,a^{-}_{\alpha}\,+\,\frac{1}{2}-\alpha\,,
\end{equation}
where
\begin{equation}
   \label{3}
a^{\pm}_{\alpha}=\frac{1}{\sqrt{2}}\left(\mp\frac{d}{dx}+x+\alpha x^{-1}\right)
\end{equation}
and $\alpha=-2$, or $\alpha=1$.

Although $a^{+}$ and $a^{-}$ are formally conjugate to each other, they cannot be considered as the increasing and the decreasing operators because of the "wrong" commutators with $H\,.$ Nevertheless, there are operators $A^{+}$ and $A^{-}$
that satisfy
\begin{equation}
   \label{4}
 \left[A^{\pm},\,H \right]=\mp 2 A^{\pm}
 \end{equation}
 increasing or decreasing the energy  by 2. Namely,
 \begin{equation}
   \label{5}
A^{+}=a^{+}_{\alpha}\,a^{+}_{-\alpha}=\frac{1}{2}\left(\frac{d^{2}}{dx^{2}}+x^{2}-2x^{-2}-2x\frac{d}{dx}-1\right)
\end{equation}
and
\begin{equation}
   \label{6}
A^{-}=a^{-}_{-\alpha}\,a^{-}_{\alpha}=\frac{1}{2}\left(\frac{d^{2}}{dx^{2}}+x^{2}-2x^{-2}+2x\frac{d}{dx}+1\right)\,.
\end{equation}

Now we can find the "vacuum" states $\Psi^{(\alpha)}_{0}(x)$ as the solutions of the differential equations
\begin{equation}
   \label{7}
a^{-}_{\alpha}\,\Psi^{(\alpha)}_{0}(x)=0
\end{equation}
on the semi-axis $ 0< x < +\infty\,.$

For $\alpha=-2$, it is \footnote{The normalizing factors of wave functions are dropped in what follows.}
\begin{equation}
   \label{8}
 \Psi^{(-2)}_{0}(x)=x^{2}\,\exp\left(-\frac{x^{2}}{2}\right)\,,
\end{equation}
with the "vacuum" energy $E^{(-2)}_{0}=\frac{5}{2}\,.$

The successive actions of operator $A^{+}$ give the states
$
\ \ \Psi^{(-2)}_{n}=\left(A^{+}\right)^{n}\Psi^{(-2)}_{0}\ \
$
with the energies $E^{(-2)}_{n}=\frac{5}{2}+2n\,.$
They have the form
$$
\Psi^{(-2)}_{n}(x)=P_{2n+2}(x)\,\exp\left(-\frac{x^{2}}{2}\right)\,,
$$
where $P_{2n+2}(x)$ is a polynomial with only even powers of $x$ and $P_{2n+2}(0)=0\,.$
Thus, for $\alpha=-2$ we get the above-mentioned regular at $x=0$ solutions that give a basis in
$L_{2}(0,+\infty)\,.$

For $\alpha=1$, the wave functions are singular at $x=0$. The solution of eq. (\ref{7}) is
\begin{equation}
   \label{9}
 \Psi^{(1)}_{0}(x)=x^{-1}\,\exp\left(-\frac{x^{2}}{2}\right)\,,
\end{equation}
with the "vacuum" energy $E^{(1)}_{0}=-\frac{1}{2}\,.$

And
\begin{equation}
   \label{10}
 \Psi^{(1)}_{n}(x)=\left(A^{+}\right)^{n}\Psi^{(1)}_{0}(x)=\left(x^{-1}+P_{2n-1}(x)\right)\,\exp\left(-\frac{x^{2}}{2}\right)\,,
\end{equation}
with $E^{(1)}_{n}=-\frac{1}{2}+2n\,.$
Here, $P_{2n-1}(x)$ is a polynomial with only odd powers of $x$.

From the equation
 \begin{equation}
   \label{11}
\lim_{\varepsilon\rightarrow 0}\frac{\int^{b}_{a}\left|\Psi^{(1)}_{n}(x)\right|^{2}dx}
{\int^{\infty}_{\varepsilon}\left|\Psi^{(1)}_{n}(x)\right|^{2}dx}=0\,, \ \ \ a>0\,,
\end{equation}
it follows that the new wave functions
  are concentrated in the infinitesimal vicinity of $x=0\,.$

In spite of the divergency of the integral
$$
\int^{+\infty}_{0}x^{-2}\,\exp(-x^{2})dx\,,
$$
it is possible to give a consistent meaning to the norm of the state $\Psi^{(1)}_{0}(x)\,.$
Note that the norms of the states  $\Psi^{(-2)}_{n}(x)$ are expressed in terms of  $\Gamma -$functions, say
$$
\left(\Psi^{(-2)}_{0}(x)\,,\ \Psi^{(-2)}_{0}(x)\right)=\int^{+\infty}_{0}x^{4}\,\exp(-x^{2})dx=\frac{1}{2}
\Gamma\left(\frac{5}{2}\right)\,.
$$
According to the definition of $\Gamma -$function at negative arguments, we regularize the divergent integral as
$$
\int^{+\infty}_{0}x^{-2}\,\exp(-x^{2})dx=\frac{1}{2}
\int^{+\infty}_{0}t^{-\frac{3}{2}}\,\exp(-t)dt=\frac{1}{2}
\Gamma\left(-\frac{1}{2}\right)
$$
and get the norm
\begin{equation}
   \label{12}
\left(\Psi^{(1)}_{0}(x)\,,\ \Psi^{(1)}_{0}(x)\right)=\frac{1}{2}
\Gamma\left(-\frac{1}{2}\right)=-\sqrt{\pi}\,.
\end{equation}

In this case, one can easily verify the orthogonality of the functions $\Psi^{(1)}_{n}$
$$
\left(\Psi^{(1)}_{n}(x)\,,\ \Psi^{(1)}_{m}(x)\right)=0\,,\ \ \ n,m=0,1,...\,,\ \ \  n\neq m\,,
$$
and the self-adjointness of the operator $H\,.$

 Although the norm of $\Psi^{(1)}_{0}$  is negative and the vacuum state (\ref{6}) is unphysical, the other functions $\Psi^{(1)}_{n}$ have positive norms
\begin{equation}
   \label{13}
\left(\Psi^{(1)}_{n}(x)\,,\ \Psi^{(1)}_{n}(x)\right)>0\,,\ \ \ n\geq 1\,,
\end{equation}
and correspond to physical states.
(The connection between singularities and indefinite metric of functional spaces was noticed in \cite{(Ber)} in other context.)

\textbf{2. The system of two quantum harmonic oscillators.}

Now, consider the quantum system of two harmonic oscillators with the Hamiltonian
\begin{equation}
   \label{14}
H=\frac{1}{2}\left(-\frac{\partial ^{2}}{\partial x^{2}}-\frac{\partial ^{2}}{\partial y^{2}}+x^{2}+y^{2}\right)=
\frac{1}{2}\left(-\frac{\partial ^{2}}{\partial z \partial \bar{z}}+\bar{z}z\right)\,.
\end{equation}
Here, as usual,
$$
z=x+\imath y\,,\ \bar{z}=x-\imath y\,,\ \frac{\partial}{\partial z}=\frac{\partial}{\partial x}-\imath\frac{\partial}{\partial y}\,,\ \frac{\partial}{\partial \bar{z}}=\frac{\partial}{\partial x}+\imath\frac{\partial}{\partial y}\,,
$$
 and
$$
\frac{\partial}{\partial z} z=\frac{\partial}{\partial \bar{z}} \bar{z}=2\,.
$$
Now, introduce two pairs of relatively conjugate operators $b^{-}_{+}=\left( b^{+}_{+}\right)^{\ast}$ and $b^{-}_{-}=\left( b^{+}_{-}\right)^{\ast}$
\begin{equation}
   \label{15}
b^{+}_{+}=\frac{1}{2}\left(-\frac{\partial}{\partial\bar{z}}+z \right),\ \
b^{-}_{+}=\frac{1}{2}\left(\frac{\partial}{\partial z}+\bar{z} \right),\ \ \ \left[b^{-}_{+},\,b^{+}_{+} \right]=1\,,
\end{equation}
\begin{equation}
   \label{16}
b^{+}_{-}=\frac{1}{2}\left(-\frac{\partial}{\partial z}+\bar{z} \right),\ \
b^{-}_{-}=\frac{1}{2}\left(\frac{\partial}{\partial\bar{z}}+z \right),\ \ \ \left[b^{-}_{-},\,b^{+}_{-} \right]=1\,.
\end{equation}
These pairs of operators act independently from each other:
\begin{equation}
   \label{17}
\left[b^{+}_{-},\,b^{+}_{+} \right]= \left[b^{-}_{-},\,b^{+}_{+} \right]=\left[b^{+}_{-},\,b^{-}_{+} \right]=\left[b^{-}_{-},\,b^{-}_{+} \right]=0\,.
\end{equation}
The Hamiltonian (\ref{14}) can be written in the form
\begin{equation}
   \label{18}
H=\frac{1}{2}\left(b^{+}_{+}b^{-}_{+}+b^{+}_{-}b^{-}_{-} \right)+1\,.
\end{equation}
It is convenient to introduce the charge operator $Q$ (it can be considered as the spin as well) 
\begin{equation}
   \label{19}
Q=\frac{1}{2}\left(-\bar{z}\frac{\partial}{\partial \bar{z}} +z\frac{\partial}{\partial z} \right)=\frac{1}{2}\left(b^{+}_{+}b^{-}_{+}-b^{+}_{-}b^{-}_{-} \right)
\end{equation}
that commutes with $H$
\begin{equation}
   \label{20}
\left[Q,\,H \right]=0\,.
\end{equation}

Due to the following commutators
\begin{equation}
   \label{21}
   \left[b^{+}_{+},\,H \right]=-b^{+}_{+},\ \ \left[b^{-}_{+},\,H \right]=b^{-}_{+},\ \ \left[b^{+}_{-},\,H \right]=-b^{+}_{-},\ \ \left[b^{-}_{-},\,H \right]=b^{-}_{-}\,,
\end{equation}
\begin{equation}
   \label{22}
\left[b^{+}_{+},\,Q \right]=-b^{+}_{+},\ \ \left[b^{-}_{+},\,Q \right]=b^{-}_{+},\ \ \left[b^{+}_{-},\,Q \right]=b^{+}_{-},\ \ \left[b^{-}_{-},\,Q \right]=-b^{-}_{-}\,,
\end{equation}
operator $b^{+}_{+}\left( b^{-}_{+}\right)$ is the creation (annihilation) operator of a particle\footnote{We call the excitation a particle having in mind its status in QFT. } with the energy $1$ and the charge $1$ and $b^{+}_{-}\left( b^{-}_{-}\right)$ is the creation (annihilation) operator of its antiparticle, that is, a particle with the energy $1$ and the charge $-1\,.$

The ordinary vacuum state
\begin{equation}
   \label{23}
\Psi_{0}=\exp(-\frac{\bar{z}z}{2} )
\end{equation}
satisfies the equations
\begin{equation}
   \label{24}
b^{-}_{+}\Psi_{0}=b^{-}_{+}\Psi_{0}=0\,.
\end{equation}
In this state, the charge $Q=0$ and the energy $E=1$ (the sum of minimal energies of two  quantum harmonic oscillators).

Other eigen-states of the operators $H$ and $Q$ can be generated from the state $\Psi_{0} $
in the usual way
$$
 \left(b^{+}_{+}\right)^{k}\left(b^{+}_{-}\right)^{n}\Psi_{0}\,.
 $$
 They are represented as crosses at the verticies of the lattice at the Figure 1.
Here, the edges of the lattice correspond to the actions of the operators $b^{+}_{+},\,b^{+}_{-}$ and $b^{-}_{+},\,b^{-}_{-}\,.$
\begin{figure}
\includegraphics[height=0.9\textwidth,angle=-90]{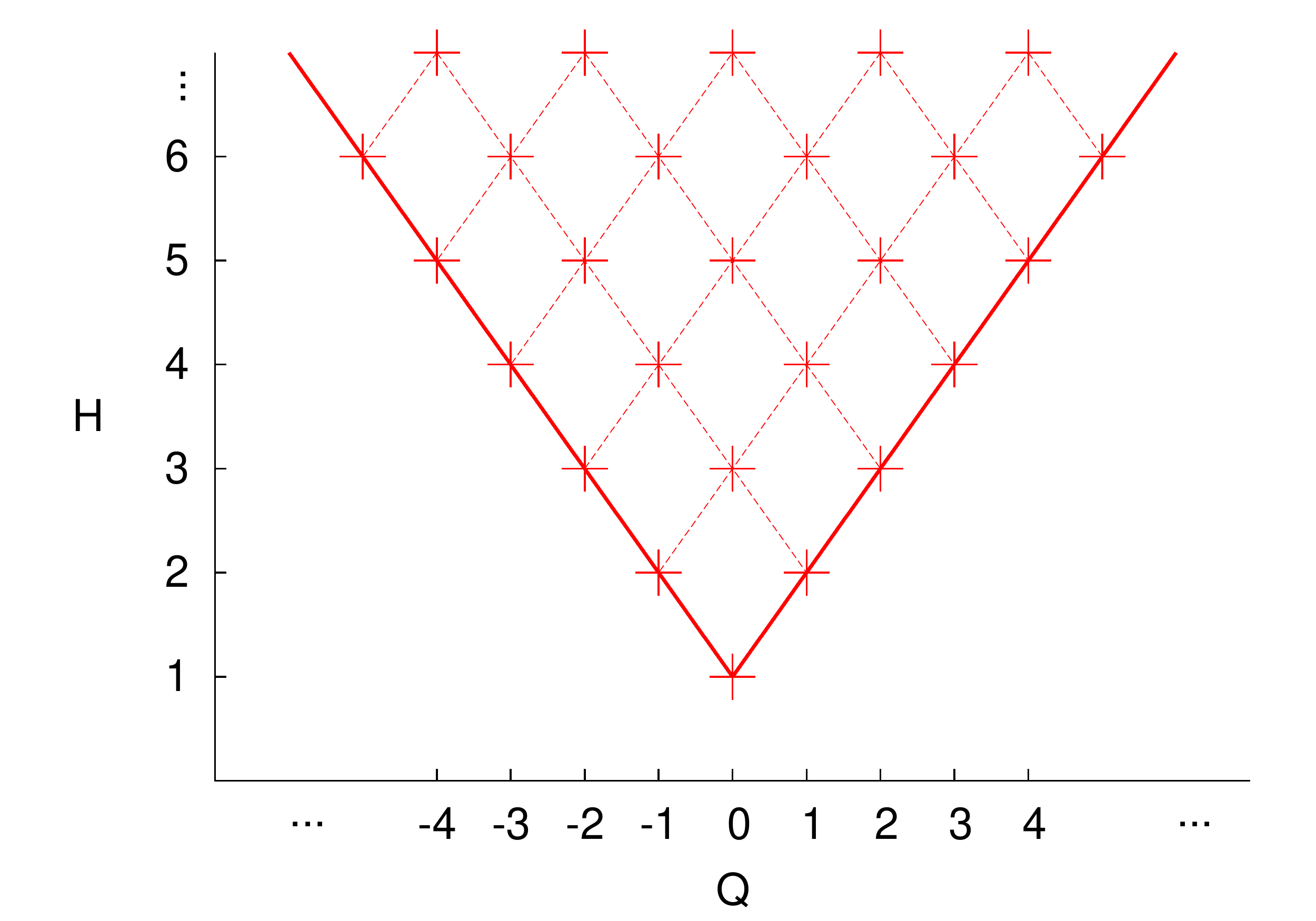}
\caption{The eigen-states of the operators $H$ and $Q$ generated from the state $\Psi_{0}\,. $}
\end{figure}

However, these states do not exhaust the set of all eigen-states  of the operators $H$ and $Q\,.$
Consider the functions
\begin{equation}
   \label{25}
 \Omega_{\lambda\mu}=\bar{z}^{\lambda}z^{\mu}\exp(-\frac{\bar{z}z}{2} )\,.
\end{equation}
They are the eigen-functions of the charge operator
\begin{equation}
   \label{26}
Q \Omega_{\lambda\mu} =(-\lambda+\mu)\Omega_{\lambda\mu}\,.
\end{equation}
Moreover, $ \Omega_{\lambda0}$ and $ \Omega_{0\mu}$ are the eigen-functions of the Hamiltonian $H$
\begin{equation}
   \label{27}
H \Omega_{\lambda0}=(\lambda+1) \Omega_{\lambda0}\,,\ \ H \Omega_{0\mu}=(\mu+1) \Omega_{0\mu}\,.
\end{equation}
They satisfy the following equations
\begin{equation}
   \label{28}
b^{-}_{+}\Omega_{\lambda0}=0\,,\ \ \ b^{-}_{-}\Omega_{0\mu}=0\,
\end{equation}
and can be considered as the states without positively charged particles $\left(\Omega_{\lambda0}\right)$ and the states without negatively charged particles $\left(\Omega_{0\mu}\right)\,.$

Now, if we permit vacuum states to have fractional charges we can consider $\left(\Omega_{\lambda0}\right)$ with a noninteger  $\lambda$ as the "initial" state. The states generated from the state $\Omega_{\frac{1}{2}0}$ by the operators
$b^{+}_{+},\,b^{+}_{-}$ and $b^{-}_{-}$ are represented at Figure 2.
\begin{figure}
\includegraphics[height=0.9\textwidth,angle=-90]{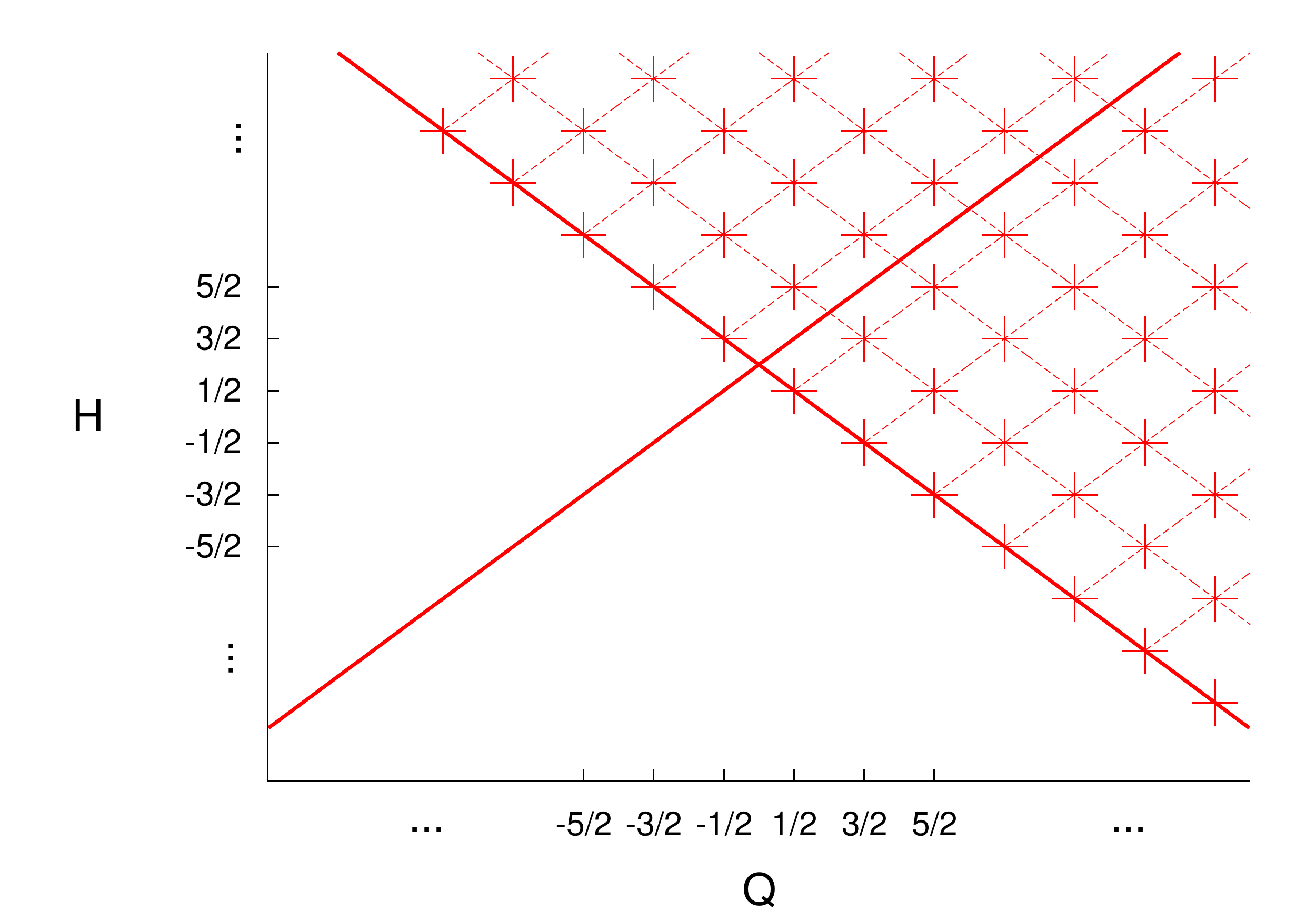}
\caption{The eigen-states of the operators $H$ and $Q$ generated from the state $\Omega_{\frac{1}{2}0}\,.$}
\end{figure}

The same class of states is obtained if one starts from other "initial" state $\Omega_{\left(\frac{1}{2}+k\right)0}\,,\ \ k=\pm1,\pm2,\ldots\,.$

   In this class of eigen-states, there is no state with minimal energy.

   The corresponding functional space has an indefinite metric. For example, the norms of the states $\Omega_{\left(-\frac{1}{2}-n\right)0}\,,\ \ n=1,2,\ldots$ have alternating signs.
 \begin{equation}
   \label{29}
\left(\Omega_{-\frac{3}{2}0}\,,\ \Omega_{-\frac{3}{2}0}\right)=\pi\Gamma\left( -\frac{1}{2}\right)=-2\pi^{\frac{3}{2}}\,,\ \
\left(\Omega_{-\frac{5}{2}0}\,,\ \Omega_{-\frac{5}{2}0}\right)>0\,,\ \ \left(\Omega_{-\frac{7}{2}0}\,,\ \Omega_{-\frac{7}{2}0}\right)<0\,,
 \end{equation}
 and so on.

 Here, the divergent integrals are regularized in the same way as in the section 1, that is,
\begin{equation}
   \label{30}
\int\int\left( \bar{z}z\right)^{-\frac{3}{2}}\exp(-\bar{z}z)d\bar{z}dz=\pi\Gamma\left( -\frac{1}{2}\right)\,,\ \ \ldots \ .
\end{equation}

Note that the states disposed at the main diagonal realize the regular representation of the Heisenberg algebra in Krein spaces studied in detail in papers \cite{(Mnats)} and \cite{(Vern)}.

The symmetric picture (Fig. 3) is obtained for eigen-states generated from $\Omega_{0\mu}\,.$
 \begin{figure}
\includegraphics[height=0.9\textwidth,angle=-90]{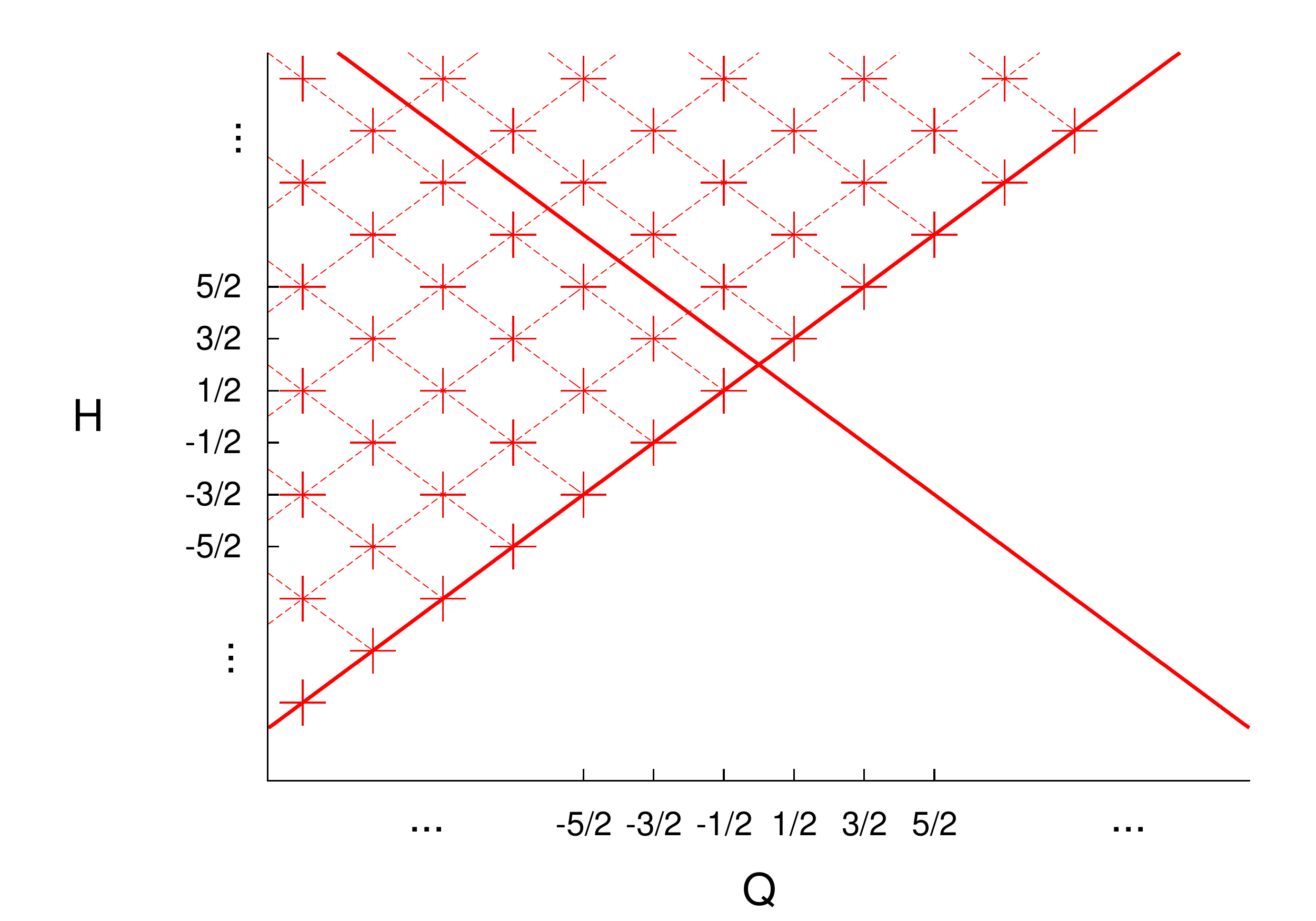}
\caption{The eigen-states of the operators $H$ and $Q$ generated from the state $\Omega_{0\frac{1}{2}}\,.$}
\end{figure}
 They have the same properties as those represented at Fig. 2.

Note that the sets of states depicted at Fig. 1, Fig.2 and Fig.3 have no common states. And all the states are orthogonal to each other.

\textbf{3. A comment concerning the singularity $x^{-2}$ of the Hamiltonian.}

       The singularities of the type $r^{-2}$ appear in Hamiltonians of multi-dimensional problems written in spherical coordinates. The reason is rather trivial: Laplacian is the differential operator of the second order.
(The singular potential $U\sim x^{-2}$ remains the simplest one in other problems as well \cite{(Fil)}.)

 If we write the Hamiltonian (\ref{14}) in polar coordinates $(r,\,\varphi)$
  \begin{equation}
   \label{31}
    -\frac{1}{2}\left(\frac{\partial^{2}}{\partial r^{2}}+\frac{1}{r}\frac{\partial}{\partial r}+\frac{1}{r^{2}}\frac{\partial^{2}}{\partial\varphi^{2}} -r^{2}\right)
  \end{equation}
    and use the substitution for wave-functions
\begin{equation}
   \label{32}
\Psi(r,\,\varphi)=r^{-\frac{1}{2}}\exp\left\{\imath\left(\frac{1}{2}+\alpha\right)\varphi\right\}\Phi(r)\,,
\end{equation}
then, for $\alpha=-2$ and $\alpha=1\,,$ we get Schr\"{o}dinger equation with the Hamiltonian considered in the section 1.

 The solutions obtained there correspond to the states with $Q=-\frac{3}{2}$ and $Q=\frac{3}{2}$ depicted at Fig. 2.

Now, the meaning of the increasing (\ref{5}) and the decreasing (\ref{6}) operators is obvious
$$
A^{+}=b^{+}_{+}b^{+}_{-}=b^{+}_{-}b^{+}_{+}\,,\ \ \
A^{-}=b^{-}_{+}b^{-}_{-}=b^{-}_{-}b^{-}_{+}\,.
$$

\textbf{4. The states generated from $\Omega_{-n0}$ and $\Omega_{0-n}\,.$ }

Now, consider the states generated from $\Omega_{-n0}\ \ (n=1, 2,\ldots )\,.$
It is convenient to consider them as the limit of the states $ \Psi_{\varepsilon}$ generated from $\Omega_{(-n+\varepsilon)0} $ that are represented by the picture similar to Fig. 2.

 Some of them tend to the states of Fig. 1 as $\varepsilon\rightarrow 0\,.$ Let us call them "ordinary" states.

The limit of the other states is represented at the right-hand side of Fig. 4.
\begin{figure}
\includegraphics[height=0.9\textwidth,angle=-90]{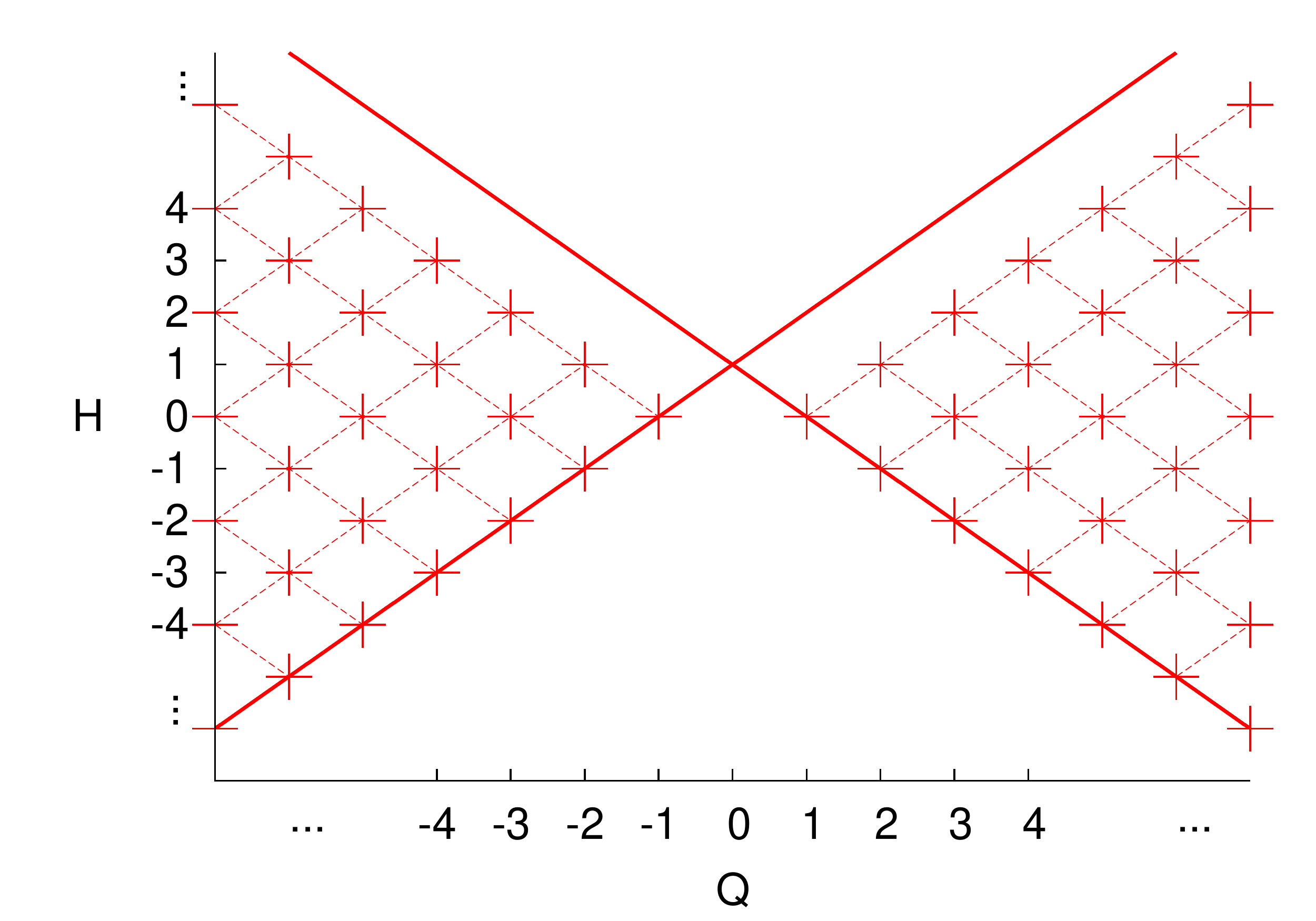}
\caption{The states generated from $\Omega_{-n0}$ and $\Omega_{0-n}\,.$}
\end{figure}

In the same way as it is done in equation (\ref{11}), one can verify that these states (not the "ordinary" ones) are localized in an infinitesimal vicinity of the origin $z=0\,,$
\begin{equation}
   \label{33}
\lim_{\varepsilon\rightarrow 0}\frac{\int\int_{a<\bar{z}z<b}\left|\Psi(\bar{z}z)\right|^{2}d\bar{z}z}
{\int\int_{\varepsilon<\bar{z}z}\left|\Psi(\bar{z}z)\right|^{2}d\bar{z}z}=0\,, \ \ \ a>0\,.
\end{equation}

Their norms defined according to (\ref{29}), (\ref{30}) tend to infinity as $\varepsilon ^{-1},.$
However,
we can renormalize the states $\Psi_{\varepsilon}\rightarrow\sqrt{\varepsilon}\Psi_{\varepsilon}\,,$
or redefine the norm of the states as
\begin{equation}
   \label{34}
\left\langle \Psi\,,\ \Psi\right\rangle=\lim_{\varepsilon\rightarrow 0}\varepsilon
\left ( \Psi_{\varepsilon}\,, \ \Psi_{\varepsilon} \right )\,.
\end{equation}
The new norm is finite for all states. For example, for the "initial" states it is
\begin{equation}
   \label{35}
\left\langle \Omega_{-n0}\,,\ \Omega_{-n0}\right\rangle=\lim_{\varepsilon\rightarrow 0}\varepsilon
\left ( \Omega_{(-n+\varepsilon)0}\,, \ \Omega_{(-n+\varepsilon)0} \right )=\pi \,res \,\Gamma(-n+1)\,.
\end{equation}

The metric of the space of states is indefinite and degenerate. The norms of the "ordinary" states are equal to zero. So, we can factorize the space of states (identify the "ordinary" states with zero).

All the same is true for the states generated from $\Omega_{0-n}\ \ (n=1, 2,\ldots )\,.$ They are represented at the left-hand side of Fig. 4.

\textbf{5. Conclusion.}

Thus, we get three different sets of states represented at Fig.1 and at both sides of Fig.4. Although the states from the different sets are formed by the same particles, they have different properties because of the different vacuum states. For example, the particles in the states from  Fig. 4 are localized at origin. 

All  the states are orthogonal to each other. And the transfer between these different sets cannot be realized as a result of any interaction of the (polynomial) form  $H_{int}=P\left(b^{+}_{+},\,b^{+}_{-},\, b^{-}_{+},\,b^{-}_{-}\right )\,.$ So, from "the point of view" of an "ordinary" state at Fig. 1,  the states at Fig. 4 prove to be "dark" ones.

It is interesting to study the effect in other two-dimensional models with more realistic interaction (e.g., \cite{(Gr)}) and in multi-dimensional models as well.

\end{document}